# Understanding Volatility Spillover relationship among G7 nations and India during COVID-19

Avik Das (Corresponding Author)
Techno India University
Avikdas.1984@yahoo.com

Dr. Devanjali Nandi Das (PhD)
Assistant Professor
Techno India University

**Abstract**

**Purpose:** In the context of a COVID pandemic in 2020-21, this paper attempts to capture the interconnectedness and volatility transmission dynamics. The nature of change in volatility spillover effects and time-varying conditional correlation among the G7 countries and India is investigated.

**Methodology:** To assess the volatility spillover effects, the bivariate BEKK and t- DCC (1,1) GARCH (1,1) models have been used. Our research shows how the dynamics of volatility spillover between India and the G7 countries shift before and during COVID-19.

**Findings:** The findings reveal that the extent of volatility spillover has altered during COVID compared to the pre-COVID environment. During this pandemic, a sharp increase in conditional correlation indicates an increase in systematic risk between countries.

**Originality:** The study contributes to a better understanding of the dynamics of volatility spillover between G7 countries and India. Asset managers and foreign corporations can use the changing spillover dynamics to improve investment decisions and implement effective hedging measures to protect their interests. Furthermore, this research will assist financial regulators in assessing market risk in the future owing to crises like as COVID-19.

## Introduction

With news of Coronavirus spreading outside of China, mostly in the United States and Europe, the S&P 500 and FTSE fell more than 3% on February 24, 2020. The CAC 40, DAX, and FTSE MIB (Stock Exchange of Italy) also plummeted more than 4%, while the FTSE MIB alone fell more than 5.5 percent. The world market had a devastating meltdown on March 9th, when Italy proclaimed full lockdown, with a drop of more than 7% in the US and Europe and more than 5% in Japan and India. The World Health Organization (WHO) proclaimed a COVID-19 pandemic on March 11th, and the S&P 500 plunged 9.99 percent the next day, while the FTSE MIB fell 18.5 percent in a single day. The DAX and CAC 40 both lost more than 13%.

In a single day, Japan's stock market lost 4.5 percent, while the Nifty 50 lost 8.6 percent. Between January and April, the G7 countries' GDP plummeted by more than 40% in just four months (Fig.1). United States stock market struck the lower circuit 4 times in 10 days in March 2020. On the 13th of March 2020, the NSE and BSE in India struck a 10% lower circuit for the first time in 12 years, and trade was halted for 45 minutes. In a single day's trading, about 30 lakh crores were lost.[1].

The study of adaptive volatility transmission among various stock markets during a crisis continues to pique the interest of academics and researchers. Spillover of volatility serves as an early warning system for major systemic danger (He, Liu, & Chen, 2018). As a result, volatility spillover analysis is critical for policymakers, foreign investors, portfolio managers, and fund managers, as it provides a broad framework for understanding market links and risk in a globally integrated world. Recently, the coronavirus pandemic 2020 has triggered an immense economic shock to all the nations. COVID-19 spread has increased human suffering resulting in ambiguity about the infectiousness, appearance, and fatality of the virus (Baker et al. (2020)). This uncertainty triggers fear in investors and high volatility in financial markets (Antonakakis et al. (2013)). The global financial market was exposed to a major sell-off during the month of March-April 2020. All major economies faced around 35% crash during this volatile period except China (Xinhua, 2020). Volatility jumps have surpassed Great depression 1929, Global Financial Crisis (GFC) 2007-08 and Spanish flu 1918-19 (Baker et al. (2020)). Historically, the international stock market crash due to COVID-19 is unprecedented. Individual stock market reactions to COVID-19 were mostly due to the severity of infection to that particular country (Zhang, Hu, & Ji, 2020). Initially when it was contained in China, the situation was stable, but the global market was in free fall when it formed two other epicentres in Europe and North America. The European market showed highest volatility jumps during the US phase though the Europe segment documented a higher death rate (Ali, Alam, & Rizvi, 2021). Whereas, total volatility spillover did not increase significantly when the pandemic was contained only in China during February 2020 (Wang, Li, & Huang (2020)).

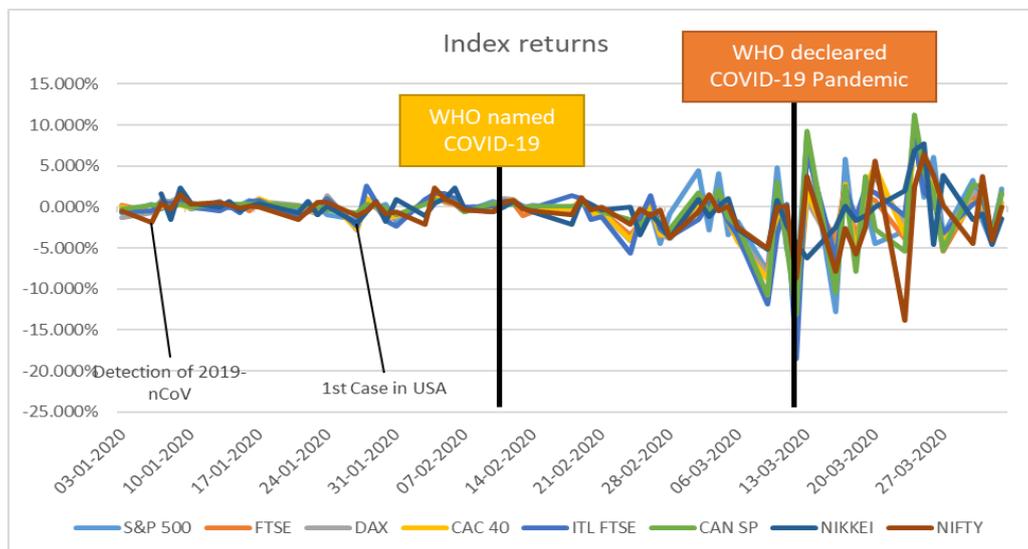

*Figure 1: Index Returns of first 3 months of 2020*

Based on the above facts, our current study focuses on the effect of volatility spillover during the second phase and third phase of COVID contamination. We are considering nature of volatility spillover effect from major economies (G7 countries) to domestic country, i.e., India during and pre-COVID situation as 1) Indian financial market exhibited massive crash due to COVID and 2) India is the still emerging market and sixth largest economy in world. Our study contributes firstly, understanding the nature of volatility spillover from G7 nations to India during and pre-COVID times. Secondly, we tried to compare the volatility spillover features between two phases that may help Indian policy makers and investors to mitigate risk in future. Thirdly, this study focuses on the volatility spillover from developed nations to the Indian market under the influence of a systematic shock that makes this study unique and innovative.

---
[1] https://www.bloombergquint.com/markets/nifty-hits-lower-circuit-for-the-first-time-since-may-2009

**Literature Review**

The earlier studies mostly focused on the connectedness of developed markets like USA, Europe and Japanese markets (Xiao & Dhesi (2010), Diebold & Yilmaz (2009), Bekaert, Ehrmann, Fratzscher, & Mehl (2014)). Some of the researchers focused on Emerging markets and Asian countries (Aggarwal, Inclan, & Leal (1999), Kumar & Mukhopadhyay (2002), Bhar & Nikolova (2007), Li & Giles (2015), Kothari, G.C., Tharugathi (2016)).

The economic impact of the COVID-19 outbreak has sparked a lot of interest, and numerous experts have looked into it from various angles. COVID-19, as a non-economic event, produced a unique financial and economic disaster (Ozili & Arun, 2020). The sudden economic shock is not only damaging but also has spillover consequences (El-Erian, 2020)[2]. Information transfer through social media and other channels has increased the contagion risk (Croce, Farroni, & Wolfskeil, 2020). Increased systemic risk and country-specific hazards in the global financial markets have a direct impact on economic intervention (Zhang, Hu, & Ji, 2020). Corbet et.al. (2021), Contessi and De Pace (2021), Akhtaruzzaman et.al. (2021) all observe volatility spillover is distinct in different stock markets. The current crude oil price volatility shock, economic policy uncertainty, and geopolitical risk are all linked by the COVID-19 epidemic ( (Sharif, Aloui, & Yarovaya, 2020), (Prabheesh & Kumar, 2021), (Rai & Garg, 2021)). Not only international markets exhibit interconnectivity, Indian sector indices also exhibits large co-movement at the onset of a crisis (Guru & Das (2021)). Lahmiri & Bekiros (2020) found exhibits of emergence of volatility connectedness among the equity, and cryptocurrency markets. In another experiment, Corbet, Larkin, & Lucey (2020) found significant volatility spillover from China to the Bitcoin market as investors took "flights to safety" measures assuming early signs of pandemic as short-term crisis.

Very limited studies have been conducted on the volatility spillover from G7 nations to Indian stock market and our study should focus about the nature and dynamics of volatility spillover during COVID times.

**Data & Methodology**

We obtain all the financial market EOD data from Yahoo Finance[3]. The time period taken for our research ranges from January 2013 through June 2021 using daily data. We subdivided our time into two sections pre-COVID from January 2013 to December 2019 and during COVID-19 from January 2020 to June 2021.

Total 1832 trading days are being considered for 8 variables. We have considered 7 major stock market indices from G7 countries which includes S&P 500 from USA, FTSE from UK, CAC40 from France, DAX from Germany, S&P/TNX Composite index from CANADA, FTSE MIB from ITALY, Nikkei 225 from Japan and Nifty 50 from India for our analysis.

---

[2] Foreign Affairs: https://www.foreignaffairs.com/articles/2020-03-17/coming-coronavirus-recession
[3] https://in.finance.yahoo.com/

In this study, we have calculated daily log returns using the following formula

$$daily\ returns = \ln(P_t - P_{t-1}) \qquad (1)$$

Then created 2 different subsets of the full sample data namely, pre-COVID (data ranging from January, 2013 to December, 2019) and during COVID (from January, 2020 till June, 2021.) It is clearly evident from Figure 1. that there exist high volatile (spikes) period during March - April 2020 throughout all the countries. Japan showcases lowest volatility spikes whereas Italy exhibits highest volatility during March – April 2020.

**Stationarity Check**

The focus of the research is on stationary time series, which are needed to drive any econometrics model. This study tests the stationarity of the daily returns of stock market indices through the unit root model introduced by Dickey and Fuller (1981) i.e., Augmented Dickey-Fuller (ADF) test and by Phillips (1987) and Phillips and Perron (1988) i.e., Phillips-Perron (PP) test and by Kwiatkowski, D., Phillips, P. C., Schmidt, P., & Shin, Y. (1992) i.e., KPSS test. The Table 1 exhibits stationarity of log return data at all three tests performed on the full sample data.

| | KPSS TEST | | ADF TEST | | PP TEST | |
|---|---|---|---|---|---|---|
| **Countries** | **KPSS Level** | **Stationarity** | **Dickey-Fuller** | **Stationarity** | **Dickey-Fuller Z(t_alpha)** | **Stationarity** |
| India | 0.14785 | stationary | -7.1314 | stationary | -25.06 | stationary |
| USA | 0.069082 | stationary | -6.124 | stationary | -30.618 | stationary |
| France | 0.084533 | stationary | -7.434 | stationary | -22.69 | stationary |
| Germany | 0.057806 | stationary | -7.1602 | stationary | -22.953 | stationary |
| UK | 0.09224 | stationary | -7.2771 | stationary | -23.023 | stationary |
| Canada | 0.075614 | stationary | -6.0175 | stationary | -28.683 | stationary |
| Italy | 0.074589 | stationary | -7.4009 | stationary | -24.456 | stationary |
| Japan | 0.068699 | stationary | -7.5257 | stationary | -21.936 | stationary |

*Table 1: Stationarity Check*

**Volatility Spillover methodology**

The stock return distribution, when compared to the standard normal distribution, is depicted with higher kurtosis value and fat tails. To address this issue, numerous stock return distributions and models have been suggested. Engle (1982) and Bollerslev (1986) created GARCH models, which are the most intriguing because these can recognise both the fat–tailed distribution and the existence of time–varying volatility. The autoregressive component of GARCH is combined with a moving average component. A mean process's delayed variance terms and delayed residual errors are included in the model. The inclusion of the moving average component in the model allows it to describe both conditional variance fluctuation across time and changes in time-dependent variance.

## BEKK GARCH

For our study we use bivariate BEKK GARCH (1991) model to capture pair wise volatility spillover among G7 and India. The following formula can be estimated for mean in BEKK:

$$r_t = \mu_t + \varepsilon_t \qquad (2)$$

$$\varepsilon_t | \Omega_{t-1} \sim N(0, H_t) \qquad (3)$$

where $r_t$ vector representing daily returns at time $t$, $\varepsilon_t$ vector representing innovation or shock for individual economy at time $t$, $\Omega_{t-1}$ representing the available market information available at time $t-1$ with its corresponding $2 \times 2$ conditional variance-covariance matrix $H_t$, and $C_t$ is vector containing the long-term coefficient drift. The conditional variance matrix can be expressed as follows:

$$H_t = C_0'C_0 + A_{1,1}'\varepsilon_{t-1}\varepsilon_{t-1}'A_{1,1} + G_{1,1}'H_{t-1}G_{1,1} \qquad (4)$$

The parameter matrix is made up of components from the lower triangular matrix C; $A_{1,1}$ matrix reveals ARCH effects and $G_{1,1}$ indicates the GARCH effects. The off-diagonal members of the parameter matrix G capture cross-market impacts, whereas the diagonal elements assess the influence of lagged volatility. The log-likelihood function of the model is expressed by

$$I_t(\theta) = -\frac{N}{2}\ln(2\pi) - \frac{1}{2}\ln(|H_t|) - \frac{1}{2}\varepsilon_t'H_t\varepsilon_t \qquad (5)$$

$L(\Theta) = \sum_{t=1}^{p} I_t(\theta)$ Where n represents the total of variables in the system, T represents the total number of observations, and v represents the unknown parameter vectors that must be estimated.

## DCC – GARCH (1,1)

We used the DCC-GARCH (Engle & Sheppard, 2001) approach to calculate the volatility spillover in our current study. The number of parameters to be estimated in the DCC-GARCH model gradually increases rather than exponentially, like in the multivariate GARCH model, effectively resolving the dimensionality issue.

Returns from n assets with expected value 0 and covariance matrix $H_t$ are assumed in the DCC model. The Dynamic Conditional Correlation (DCC-) GARCH model is thus described as follows:

$$r_t | \Omega_{t-1} \sim N(0, H_t) \qquad (6)$$

The conditional variance of each return is calculated using the residuals of the mean equation:

$$h_{i,j}^2 = \alpha_0 + \sum_{j=1}^{p_i} \alpha_j \varepsilon_{i,t-j}^2 + \sum_{j=1}^{q_i} \beta_j \sigma_{i,t-j}^2 \qquad (7)$$

where, $\sum_{j=1}^{p_i} \alpha_j + \sum_{j=1}^{q_i} \beta_j = 1$

The multimodal conditional variance $H_t$ is then calculated:

$$H_t = D_t R_t D_t \qquad (8)$$

The univariate GARCH parameters yield a (k x k) diagonal matrix of time changing standard deviations, which is represented by $D_t$.

The elements of $H_t = D_t R_t D_t$ is:

$$[H_t]_{ij} = \sqrt{h_{it} h_{jt} \rho_{ij}} \qquad (9)$$

where $\rho_{ii} = 1$

The parameters of the DCC model are estimated using the likelihood of this estimator and can be written as:

$$L = -\frac{1}{2} \sum_{t=1}^{T} (n \log(2\pi) + 2 \log|D_t| + \log|R_t| + \eta_t' R_t^{-1} \eta_t) \qquad (10)$$

Financial time series data has an own personality. For starters, it has a higher centre peak and plump tails. They also show volatility clustering. As a result, the DCC-GARCH procedure's normalcy assumptions may be broken. To avoid this issue, we used the t-DCC-GARCH technique, which applies the DCC model under the premise that market yields reflect a multivariate t-distribution, as proposed by Pesaran & Pesaran (2007).

We have employed t-statistics for comparing the DCCs between pre-COVID and during COVID periods.

$$t = \frac{Avg.DCC^h - Avg.DCC^l}{\sqrt{[\frac{\sigma_h^2}{N^h} + \frac{\sigma_l^2}{N^l}]}} \qquad (11)$$

Here, $Avg.DCC$ represent the average DCC value over high and low volatility periods. $N$ represents the number of observations and $\sigma_h^2, \sigma_l^2$ denotes the corresponding standard deviation for high and low volatility periods.

**Empirical Results**

**BEKK Analysis**

In Eq. 4, the matrices A and G are important for evaluating the relationship in the context of volatility. As shown in the Table-5, $G_{1,1}$ and $G_{2,2}$ are statistically significant at 5% level in pre-COVID and during COVID period which implies a strong GARCH (1,1) process exists in two different periods. In other words, all the countries are sensitive to their own volatility shocks.

The off-diagonal elements in the G matrix represents cross-market effect (volatility spillover) among all markets. Across all countries, except UK/IND, all reported estimates are significant at 5%. So, we can claim that except between the UK and India there exists bilateral volatility spillover between all G7 nation and India. Second, volatility spillover effects really aren't symmetrical, as research shows that the United States market has a higher impact on the Indian market than any other market. In specifically, the level of volatility transmission between the United States and India in the pre-COVID period is indicated by $G_{2,1}$ (0.3299) and $G_{1,2}$ (0.4704) in the USA/IND pairwise. Volatility transmission from the United States to India is 47.33 percent, implying that a 1% spike in returns in the United States will result in 47.33 percent volatility in India. When it comes to other countries' volatility transmission, Japan (27.79 percent) and Germany (22.48 percent) are at the top, followed by the United States. Rest of the nation's transmit roughly 15% and below in pre-COVID period towards India.

**[Insert Table 4]**

During the COVID period the transmitting nation's list got reformed. During COVID only USA, France, Germany and Japan transmitted volatility towards India and vice-versa (bi-directional). Whereas, India transmitted volatility towards Canada and Italy (Uni-directional). During COVID most of the nations exhibited higher persistence of their own volatility. In the case of USA/IND in both the periods, India transmitted greater volatility towards the USA than any other G7 countries. USA was the highest volatility transmitter (38.70%) towards India during pre-COVID period. Japan transmitted 29.76% and Germany transmitted around 22.48%. There was no change in the volatility spillover magnitude from Germany to India in two different periods.

**[Insert Table 5]**

**t-DCC Analysis**

We studied sensitivity and persistence through the t-DCC-GARCH model. This allows us to measure the real impact of G7 index price's volatility on the Indian volatility. DCC alpha ($DCC_\alpha$) and DCC beta ($DCC_\beta$) reveals how the correlations are evolving over time in an autoregressive manner. $DCC_\alpha$ provides the contribution of the realized correlation matrix from last period while $DCC_\beta$ provides the contribution of correlation matrix that is due to all previous periods. Our results pass through several diagnostic tests. The Hosking (1980) test finds no indication of serial correlation, and the Li and McLeod (1981) test finds no evidence of model misspecification. The sum of two coefficient DCC alpha ($DCC_\alpha$) and DCC beta ($DCC_\beta$) is less than 1 ($DCC_\alpha + DCC_\beta < 1$).

For further investigation we extracted the conditional correlations of all G7 nations with India and performed comparative analysis (t-statistics). The DCC estimates for India and the G7 nations index returns are greater during the COVID period than during the pre-COVID period. The DCC difference between a crisis and a non-crisis era differs per country. According to the findings, Italy has the largest difference in mean DCC (0.2442).

| Parameters | USA | France | Germany | UK | Canada | Italy | Japan |
|---|---|---|---|---|---|---|---|
| pre-COVID mean | 0.2617 | 0.3700 | 0.3737 | 0.3642 | 0.2474 | 0.3024 | 0.3344 |
| during-COVID mean | 0.3713 | 0.5582 | 0.5655 | 0.5316 | 0.4242 | 0.5466 | 0.4523 |
| mean difference | 0.1096 | 0.1883 | 0.1917 | 0.1674 | 0.1768 | **0.2442** | 0.1179 |
| t-statistic | -54.0106 | -97.3186 | -100.5279 | -86.1550 | -82.3165 | -123.3843 | -57.1103 |

*Table 2: DCC mean differences and t-statistics*

A significant jump in conditional correlation has been observed in European nations where all countries' conditional correlation jumped to more than 0.70 (max values) where Japan and USA conditional correlation remain stable with max values of 0.59 and 0.56 respectively. As Table 6. exhibits significant DCC mean difference between two periods and also statistically significant. Higher DCCs during the crisis time are consistent with findings from research on the global economic crisis (Akhtaruzzaman, Shamsuddin, & Easton (2014); Dimitriou, Kenourgios, & Simos (2013); Kim, Kim, & Lee (2015)). The dynamic conditional correlations implied in the DCC model are plotted in Table 3 (pre-COVID) and (during COVID). These figures represent time-varying patterns in correlation dynamic paths. Both periods clearly showcase higher correlation between European and Japanese market to India as compared to USA and Canada.

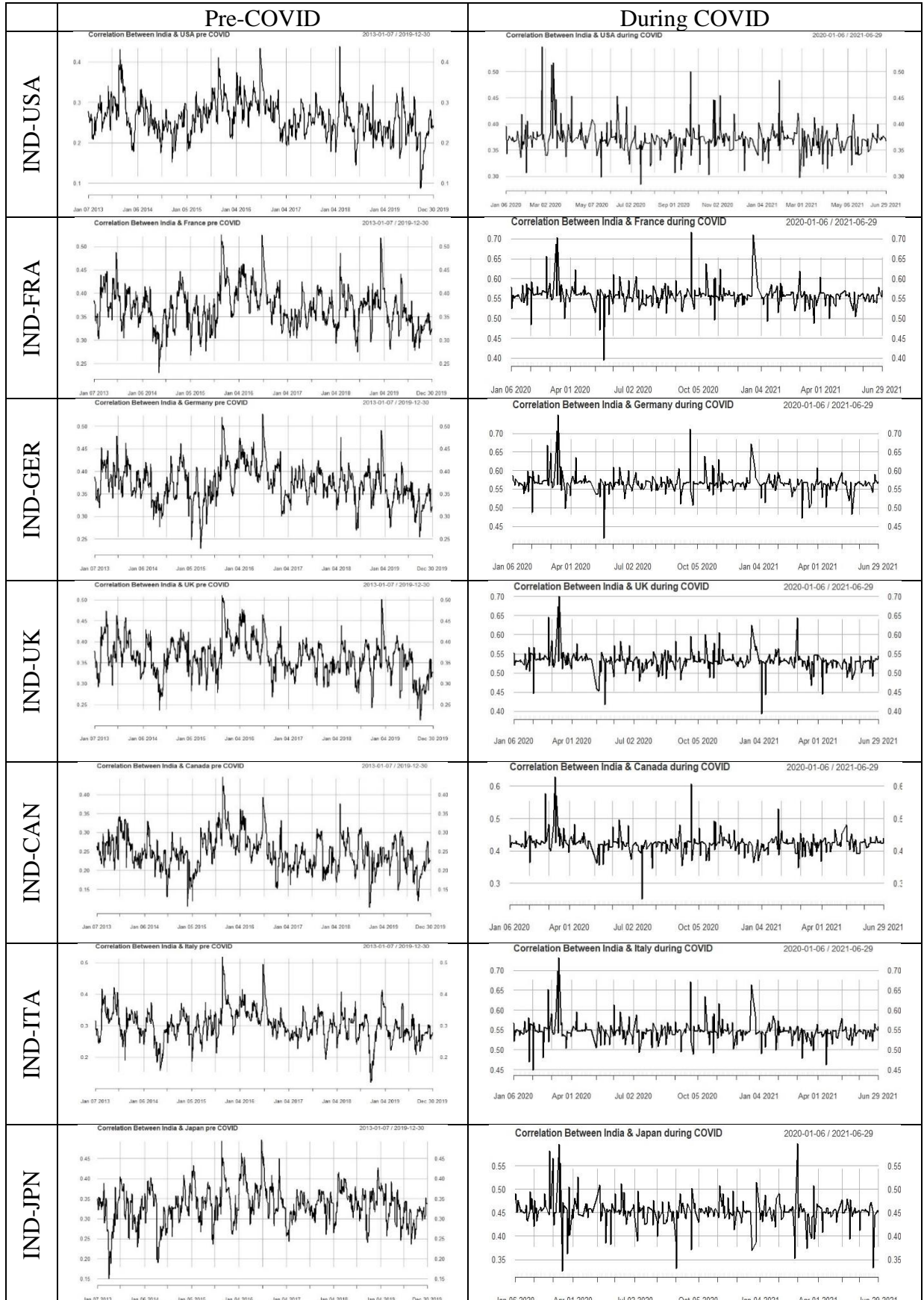

*Table 3: Comparison of Conditional Correlation pre & post COVID*

**Conclusion and Discussion**

This research examines the impact of a COVID-19 pandemic outbreak in India and the G7 countries on stock market movements in 2020. It compares volatility spillover across pre-COVID-19 and post-COVID-19 eras. During COVID, the financial markets responded significantly and in an unforeseen fashion. During the research period, the volatility spillover between India and the G7 countries increased dramatically, according to our findings.

According to the BEKK study, all G7 countries contributed significantly to India's volatility prior to COVID, although the quantity of spillover differs during COVID periods. Despite the fact that Japan and Germany were never the epicentre of COVID-19, Japan, and Germany along with the United States, continue to be the principal volatility transmitters to India. Canada and Italy are the recipients of volatility spillover from the rest of the world. During COVID times, India has become a key transmitter. In both eras, there is no major volatility spillover from the United Kingdom to India. During the COVID–19 era, dynamic conditional correlations (DCCs) between India and G7 nation index returns increased dramatically. Italy has the biggest DCC difference (0.2442), while the United States has the lowest (0.1096). From the t-stat it is evident that change in DCC means is significant in 5% significance level i.e., due to COVID-19 a major volatility spillover has been observed. Increased conditional correlation between countries increases the co-riskiness among them.

Finally, in the absence of volatility spillover, overseas investors may find that investing in India and the United Kingdom is a good way to diversify their portfolio. Due to the COVID problem, the largest transmitters of volatility are the United States, Germany, and France, thus fund managers and investors should hedge their investments against a possible catastrophic fall in these markets. COVID–19 research is still in its early stages, and we are still concerned about a third wave of infection in India. More research on the subject is needed, especially now that longer time periods are available under the COVID–19.

**Annexure**

*Table 4: BEKK Estimation of pre-COVID period (\*\*\*, \*\*, and \* significant at 0.1%, 1%, and 5%, respectively.)*

| Parameters | USA/IND Estimate | Std. Er | FRA/IND Estimate | Std. Er | GER/IND Estimate | Std. Er | UK/IND Estimate | Std. Er |
|---|---|---|---|---|---|---|---|---|
| $c_{1,1}$ | 0.003008432** | 0.0010 | 0.003008432*** | 0.0007 | 0.003008432*** | 0.0008 | 0.003008432*** | 0.0008 |
| $c_{1,2}$ | 0.001145384 | 0.0008 | 0.001632862* | 0.0008 | 0.001524113 | 0.0010 | 0.001380072* | 0.0006 |
| $c_{2,2}$ | 0.003019523*** | 0.0006 | 0.002650628*** | 0.0006 | 0.002770487*** | 0.0007 | 0.002419041*** | 0.0005 |
| $A_{1,1}$ | 0.258891137*** | 0.0536 | 0.315095539*** | 0.0539 | 0.354016614*** | 0.0775 | 0.322117606*** | 0.0667 |
| $A_{2,1}$ | -0.026147877 | 0.0552 | -0.059243352 | 0.0734 | -0.092951056 | 0.0735 | -0.031209202 | 0.0506 |
| $A_{1,2}$ | 0.43279327*** | 0.0767 | 0.275038655*** | 0.0501 | 0.201492641*** | 0.0493 | 0.180972237*** | 0.0522 |
| $A_{2,2}$ | 0.473366959*** | 0.0682 | 0.317675073*** | 0.0543 | 0.233785341** | 0.0823 | 0.335278616*** | 0.0436 |
| $G_{1,1}$ | 0.803250125*** | 0.0707 | 0.914747444*** | 0.0376 | 0.905675872*** | 0.0524 | 0.904083395*** | 0.0366 |
| $G_{2,1}$ | 0.329927316*** | 0.0473 | 0.219664054*** | 0.0505 | 0.20919569*** | 0.0592 | 0.038314383 | 0.0373 |
| $G_{1,2}$ | -0.470484718*** | 0.0424 | -0.210912254*** | 0.0332 | -0.13925401** | 0.0515 | -0.081274107** | 0.0292 |
| $G_{2,2}$ | 0.701526917*** | 0.0395 | 0.814536317*** | 0.0361 | 0.85083293*** | 0.0340 | 0.90283431*** | 0.0236 |

| Parameters | CAN/IND Estimate | Std. Error | ITA/IND Estimate | Std. Error | JPN/IND Estimate | Std. Error |
|---|---|---|---|---|---|---|
| $c_{1,1}$ | 0.003982276*** | 0.0008 | 0.003008432*** | 0.0008 | 0.003008432*** | 0.0007 |
| $c_{1,2}$ | 0.001415857** | 0.0005 | 0.001581865. | 0.0009 | 0.001243631 | 0.0012 |
| $c_{2,2}$ | 0.002633756*** | 0.0004 | 0.003026094*** | 0.0005 | 0.002381217* | 0.0012 |
| $A_{1,1}$ | 0.341117981*** | 0.0546 | 0.342375089*** | 0.0711 | 0.327233888*** | 0.0571 |
| $A_{2,1}$ | 0.037428238 | 0.0361 | 0.052047456 | 0.0780 | -0.114646081* | 0.0563 |
| $A_{1,2}$ | -0.42377285*** | 0.1040 | 0.187138954*** | 0.0482 | 0.130407257** | 0.0462 |
| $A_{2,2}$ | 0.633300619*** | 0.0715 | 0.314948668*** | 0.0442 | 0.000001 | 0.0004 |
| $G_{1,1}$ | 0.644862057*** | 0.0725 | 0.911058361*** | 0.0421 | 0.938243607*** | 0.0354 |
| $G_{2,1}$ | -0.15106348*** | 0.0458 | 0.132288295** | 0.0442 | 0.277939437*** | 0.0342 |
| $G_{1,2}$ | 0.5*** | 0.0829 | -0.14123193*** | 0.0405 | -0.2246565*** | 0.0360 |
| $G_{2,2}$ | 0.757706174*** | 0.0518 | 0.867503834*** | 0.0253 | 0.820353031*** | 0.0411 |

*Table 5: BEKK Estimates during COVID (\*\*\*, \*\*, and \* significant at 0.1%, 1%, and 5%, respectively.)*

| Parameters | USA/IND Estimate | Std. Error | FRA/IND Estimate | Std. Error | GER/IND Estimate | Std. Error | UK/IND Estimate | Std. Error |
|---|---|---|---|---|---|---|---|---|
| $c_{1,1}$ | 0.003694137* | 0.0006 | 0.004175587*** | 0.0009 | 0.003671939*** | 0.0009 | 0.0036719392* | 0.0016 |
| $c_{1,2}$ | 0.001719136* | 0.0000 | 0.003167194. | 0.0016 | 0.002885583 | 0.0022 | 0.0020157405 | 0.0013 |
| $c_{2,2}$ | 0.004215393** | 0.0013 | 0.003529443** | 0.0012 | 0.00392727* | 0.0016 | 0.0030185202 | 0.0023 |
| $A_{1,1}$ | 0.019746097 | 0.0289 | 0.18012659* | 0.0774 | 0.158539017* | 0.0751 | 0.2470329525. | 0.1433 |
| $A_{2,1}$ | 0.145895265 | 0.0914 | -0.189963473* | 0.0953 | -0.249533231* | 0.1028 | -0.0925393782 | 0.1692 |
| $A_{1,2}$ | 0.467501509*** | 0.0728 | 0.359617642*** | 0.0704 | 0.300423086*** | 0.0684 | 0.2097857342* | 0.0960 |
| $A_{2,2}$ | 0.382463782* | 0.1767 | 0.448028187*** | 0.0733 | 0.324338925*** | 0.0896 | 0.3270248831*** | 0.0799 |
| $G_{1,1}$ | 0.816014404*** | 0.0537 | 0.882229863*** | 0.0513 | 0.918140193*** | 0.0512 | 0.9135387939*** | 0.0816 |
| $G_{2,1}$ | 0.387074376*** | 0.0927 | 0.159209316* | 0.0733 | 0.224867712* | 0.0882 | 0.0673314349 | 0.1087 |
| $G_{1,2}$ | -0.410187787*** | 0.0961 | -0.157345795*** | 0.0413 | -0.123927943** | 0.0385 | -0.0719670735 | 0.0625 |
| $G_{2,2}$ | 0.695365878*** | 0.0495 | 0.815237089*** | 0.0409 | 0.823308244*** | 0.0565 | 0.9004966874*** | 0.0845 |

| Parameters | CAN/IND Estimate | Std. Error | ITA/IND Estimate | Std. Error | JPN/IND Estimate | Std. Error |
|---|---|---|---|---|---|---|
| $c_{1,1}$ | 0.0036719392* | 0.0014 | 0.004239857*** | 0.0013 | 0.003671939*** | 0.0010 |
| $c_{1,2}$ | 0.0022259774. | 0.0013 | 0.004781541** | 0.0018 | 0.00111851 | 0.0040 |
| $c_{2,2}$ | 0.003743225*** | 0.0007 | 0.0038038** | 0.0012 | 0.003634052** | 0.0014 |
| $A_{1,1}$ | 0.0935991218 | 0.0998 | 0.316466282*** | 0.0836 | 0.135648503 | 0.1877 |
| $A_{2,1}$ | 0.0370543079 | 0.0749 | 0.114217036 | 0.1048 | -0.239056963* | 0.1028 |
| $A_{1,2}$ | 0.4610281887*** | 0.1159 | 0.25105008** | 0.0774 | 0.220029942** | 0.0704 |
| $A_{2,2}$ | 0.9738703314*** | 0.1021 | 0.479705292*** | 0.0934 | 0.000001 | 0.0004 |
| $G_{1,1}$ | 0.9518503661*** | 0.0531 | 0.905068849*** | 0.0656 | 0.996078207*** | 0.0735 |
| $G_{2,1}$ | 0.0721256943 | 0.0764 | 0.063710718 | 0.0814 | 0.297606848*** | 0.0493 |
| $G_{1,2}$ | -0.377851072*** | 0.0562 | -0.164878615** | 0.0600 | -0.237451651. | 0.1214 |
| $G_{2,2}$ | 0.4852358616*** | 0.0880 | 0.802024195*** | 0.0658 | 0.724263815*** | 0.0810 |